\documentstyle[aps,preprint]{revtex}
\begin{document}
\newcommand{\be}{\begin{equation}}
\newcommand{\ee}{\end{equation}}
\newcommand{\bds}{\begin{description}}
\newcommand{\eds}{\end{description}}
\newcommand{\bfl}{\begin{flushleft}}
\newcommand{\efl}{\end{flushleft}}
\newcommand{\ud}{\underline}
\title{Effects of Impurities in Random Sequential Adsorption 
on a One-Dimensional Substrate}
\author{Jae Woo Lee\footnote{ E-mail:jwlee@craft.camp.clarkson.edu} }
\address{Department of Physics, Inha University, Inchon 402-751, KOREA
\\
Department of Physics, Clarkson University, 
Potsdam, New York 13699-5820 USA}

\maketitle

\begin{abstract}
We have solved the kinetics of random sequential adsorption of linear $k$-mers
on a one-dimensional disordered substrate for the random sequential
adsorption initial condition and for the random initial condition.
The jamming limits $\theta(\infty, k^{'}, k)$ at fixed length of linear
$k$-mers
have a minimum point at a particular density of the
linear  $k^{'}$-mers impurity for both cases.
The coverage of the surface and the jamming limits 
are compared to the
results for Monte Carlo simulation. The Monte Carlo results 
for the jamming limits are in good agreement
with the analytical results. The continuum limits are derived from 
the analytical results on
lattice substrates.
\end{abstract}
\pacs{05.70.Ln,68.10.Jy}
%
\newpage
\narrowtext

Random sequential adsorption (RSA) of linear $k$-mers on a lattice is a model of
nonequilibrium
deposition process\cite{Evans,BP,Privman}.
The linear $k$-mers are deposited at random, sequentially
and irreversibly on a substrate without diffusion and detachment. The 
incoming particles do not overlap previously deposited particles. The 
adhesion of colloidal particles to solid substrate serves as an experimental 
realizations of RSA\cite{Feder,Onoda}. The surface coverages converge to the jamming
limits at long times. RSA of linear $k$-mers on a one-dimensional 
lattice has been exactly solved by various methods\cite{Flory,Cohen}.
The kinetics of RSA
on a one-dimensional disordered substrates occupied with point impurities
has been studied numerically by Milo\u{s}evi\'{c}
 and \u{S}vrati\'{c}\cite{MS} and solved analytically
by Ben-Naim and Krapivsky\cite{BK}. Recently, the kinetics of RSA on 
a two-dimensional disordered substrata with point impurities
has been studied by Lee\cite{Lee}
using Monte Carlo method.

In the present work we have studied the RSA of linear $k$-mers on a one-dimensional
disordered substrate for the random sequential adsorption initial
condition and for the random initial condition.

Let the initial density of $k^{'}$-mer impurities be $\rho_o$. Initially,
$k^{'}$-mer impurities of density $\rho_o$ are adsorbed randomly and 
sequentially on an empty one-dimensional substrate.
Consider the elapsed time $t_o$ at
which the density of $k^{'}$-mer impurities is $\rho_o$. Let $P_m(t;k^{'})$
denote the probability that $m$-consecutive sites are empty.
The $k^{'}$-mers are adsorbed on a clean surface. The rate equations for 
these probabilities are\cite{Evans,BP,BK}
\begin{eqnarray}
\frac{d P_m(t;k^{'})}{dt} & = &
-(k^{'} -m +1) P_{k^{'}}(t;k^{'}) -2 \sum_{j=1}^{m-1} P_{k^{'}+j}(t
;k^{'}),
~~m \leq k^{'} \\
&=&
-(m-k^{'}+1) P_m(t;k^{'}) -2 \sum_{j=1}^{k^{'}-1} P_{m+j}(t;k^{'}),
~~m \geq k^{'} 
\end{eqnarray}
The first term of the right-hand side corresponds to the $k^{'}$-mer
covering fully the $m$-site sequence ($m \leq k^{'}$) or filling with 
it ($m \geq k^{'}$). The second term describes the probabilities of  
deposition  events in which the $m$-site sequence is made non-empty by
a partial overlap by the incoming $k^{'}$-mer. Put the trial solution  
$P_m(t;k^{'})$ as
\be
P_m(t;k^{'} \leq m) = a(t;k^{'}) e^{-mt}
\ee
where
\be
a(t;k^{'}) = \exp \left[ (k^{'}-1)t -2 \sum_{j=1}^{k^{'}-1} \frac{1-e^{-jt}}
{j} \right] 
\ee
The coverage by $k^{'}$-mers is given by
\begin{eqnarray}
\theta(t;k^{'}) & = & 1-P_1(t,k^{'}) \\
 & = & k^{'} \int_0^{t} du \exp \left[ -u -2 \sum_{j=1}^{k^{'}-1}
 \frac{1-e^{-ju}}{j} \right]
\end{eqnarray}
The elapsed time $t_o$ is defined as the time that 
the coverages of the surface reaches the initial density of 
the impurities, $\rho_o$: 
\be
\rho_o = k^{'} \int_0^{t_o} du \exp \left[ -u -2 \sum_{j=1}^{k^{'}-1}
 \frac{1-e^{-ju}}{j} \right]
\ee
If $k^{'}=1$, then $t_o = - \ln (1-\rho_o)$. Therefore,
the probability $P_m (t_o,k^{'}=1)$ is given by
\be
P_m (t_o,k^{'}=1) = (1-\rho_o)^m
\ee
This result is consistent with the previous result of Ben-Naim and
Krapivsky\cite{BK}.
When $k^{'}=2$, $t_o = -\ln [ 1+ \frac{1}{2} \ln (1-\rho_o) ]$ and
\be
P_m(t_o,k^{'}=2) = e^{-mt_o} a(t_o;k^{'}=2)
\ee
The 
probability $P_m(t;k^{'},k)$ for adsorption of a $k$-mer
on a substrate occupied initially
with $k^{'}$-mer
impurities of density $\rho_o$, 
follows the same rate equations of Eq.(1)
and Eq.(2) with $k^{'}$ replaced by $k$.
Let us the initial density of impurities is $\rho_o$ and
consider the trial solution for $P_m (t;k^{'},k)$ for $m \ge k$,
\be
P_m (t;k^{'},k) =P_m(0) f_m(t) e^{-mt}
\ee
where $f_m(0)=1$ and $P_m(0)=P_m(t_o;k^{'})$. 
Next we substitute Eq.(10) to Eq.(2) and
solve for $f_m(t)$.  We obtain $f(t)$ as
\be
f_m(t) = \exp \left[ (k-1)t 
-2 \sum_{j=1}^{k-1} \frac{(1-e^{-jt})}{j} \frac{P_{m+j}(0)}{P_m(0)}
 \right]
\ee
Therefore, the probability $P_m(t;k^{'},k) $ is given by
\be
P_m(t;k^{'},k) =a(t_o;k^{'}) e^{-mt_o} \exp \left[ -(m-k+1)t 
-2 \sum_{j=1}^{k-1} \frac{(1-e^{-jt})}{j} \frac{P_{m+j}(0)}{P_m(0)}
 \right]
\ee
If $k \ge k^{'}$, then $\frac{P_{m+j}(0)}{P_m(0)}=e^{-jt_o}$.
 From Eq.(2) the rate equation for $P_1(t;k^{'},k)$ is given by
\be
\frac{d P_1(t;k^{'},k)}{dt} = - k P_k(t;k^{'},k)
\ee
Solution of $P_1(t;k^{'},k)$ for $k \ge k^{'}$ is obtained as
\be
P_1(t;k^{'},k) = P_1(0)-k a(t_o;k^{'}) e^{-kt_o}
\int_0^t du  \exp \left[ -u
-2 \sum_{j=1}^{k-1} \frac{(1-e^{-ju})}{j} e^{-jt_o} \right]
\ee
where
\be 
P_1(0)=1-k^{'} \int_0^{t_o} du \exp \left[ -u
-2 \sum_{j=1}^{k^{'}-1} \frac{1-e^{-ju}}{j} \right]
\ee
and
\be
a(t_o;k^{'}) = \exp \left[ (k^{'}-1) t_o -2 \sum_{j=1}^{k^{'}-1}
\frac{1-e^{-jt_o}}{j} \right]
\ee
The coverage for $k \ge k^{'}$ is obtained as
\begin{eqnarray}
\theta(t;k^{'},k) &=& 1-P_1(t;k^{'},k) \\
&=& \rho_o +k a(t_o;k^{'}) e^{-kt_o} \int_0^{t} du
\exp \left[ -u
-2 \sum_{j=1}^{k-1} \frac{(1-e^{-ju})}{j} e^{-jt_o} \right]
\end{eqnarray}
For $k < k^{'}$ the initial probability $P_m(0)$ is obtained from
Eq.(1) as 
\be
P_m (0) = 1-\int^{1}_{e^{-t_o}} dv \left\{ (k^{'}-m+1)+2
\sum_{j=1}^{m-1} v^j \right\} \exp \left[ -2 \sum_{j=1}^{k^{'}-1} 
\frac{(1-v^j)}{j} \right]
\ee
where $v=\exp(-t)$.
Substituting $P_m(0)$ in Eq.(12) we obtain $P_m(t;k^{'},k)$.
Integrating Eq.(13) we calculate the coverage $\theta(t;k^{'},k)$ 
for $k < k^{'}$.
When $k^{'}=1$, $t_o = - \ln ( 1- \rho_o)$,  
and $a(t_o, k^{'}=1) = 1$. We substitute these values into
Eq.(18). The coverage of surface occupied initially 
with point impurities then follows as
\be
\theta(t; k^{'}=1,k) = \rho_o +k(1-\rho_o)^k \int^t_0 du 
\exp \left[ -u -2 \sum^{k-1}_{j=1} \frac{1-e^{-ju}}{j} (1-\rho_o)^j 
\right]
\ee
These results are consistent with 
results of Ben-Naim and Krapivsky\cite{BK}.
The jamming limit for the dimer deposition ($k=2$) is $\theta(
\infty, k^{'}=1, k=2)=1-(1-\rho_o) \exp[-2(1-\rho_o)]$. The jamming
limits have a minimum value $\theta_{\rm min}(\infty, k^{'}=1, k=2)
=1-e^{-1}/2=0.8160$ \ldots at $\rho_o =1/2$.
When $k^{'}=2$, $t_o=-\ln [ 1+ \frac{1}{2} \ln (1- \rho_o)]$, 
and $ a(t_o, k^{'}=2) = (1-\rho_o) / [ 1 +
\frac{1}{2} \ln(1-\rho_o)]$. Substituting these values into Eq. (18) we
obtain coverages for $k^{'} =2$ as 
\be 
\theta(t;k^{'}=2,k) = \rho_o+ k (1-\rho_o) e^{-(k-1)t_o}
 \int^t_0 \exp \left[ -u -2 \sum^{k-1}_{j=1}
\frac{(1-e^{-ju})}{j} e^{-jt_o}
\right] du
\ee
For ($k^{'}=2, k=1$) the jamming limit is trivially obtained as 
$\theta(\infty;k^{'}=2, k=1) =1$. For the deposition of dimer olny
 $(k^{'}=2, k=2)$, the jamming limit is consistent with the previous
results as $\theta(\infty; k^{'}=2, k=2) =1-e^{-2}=0.8646
 \cdots$\cite{Flory,Cohen}.
For $(k^{'} =2, k \geq 3) $, we obtain the jamming limits by integrating
Eq. (21).
The initial elapsed time $t_o$ is numerically calculated
 by using Eq.(7) when $k^{'} > 2$. Using the
time $t_o$, we calculate the coverage from Eq.(18). 
The jamming limits $\theta(t=\infty; k^{'}=2, k)$ are plotted
in Fig.1. The solid lines in Fig.1 represent  
results obtained by numerical integration of Eq.(18). 
The symbols in Fig.1 represent the Monte Carlo results
for a one-dimensional lattice of size $L=10^5$ using periodic boundary
conditions and $10^3$ configurational averages.
The Monte Carlo results are in good agreement with the results
of numerical calculations. 
The appearance of the minimum point of the jamming limits is consistent
with the previous exact results\cite{BK} and Monte Carlo 
results \cite{MS,Lee} on the adsorption of $k$-mers on a substrate
 occupied 
initially with point impurities. At low densities of $k^{'}$-mer
impurities the jamming limits decrease with increasing $\rho_o$. 
In this regime effects of impurities are to reduce the available space
for $k$-mers as compared to the empty substrate. However, at high
densities of $k^{'}$-mer these quenched impurities are already
close to the jamming state. So that only a small fraction of $k$-mers 
is
adsorbed on the substrate. The minimum point of the jamming limits
decreases with increasing length of the $k$-mers. 

Another simple solvable case is when the impurities are distributed
randomly, i.e. $P_m (0) = \lambda^m $ with $ \lambda = 
[ 1+ k^{\prime-1} \rho_o (1-\rho_o)^{-1} ]^{-1}$.
In this case, we obtain as $t_o=0$, $P_m (0)=\lambda^m$ and
$P_{m+j}(0)/P_m(0)=\lambda^j$. The coverage is obtained as 
\be
\theta(t;k^{\prime},k)=1-\lambda +k \lambda^k \int^t_0
du \exp \left[ -u -2 \sum^{k-1}_{j=1} \frac{1-e^{-ju}}{j}
\lambda^j \right]
\ee
When $k=2$, the jamming limit is given as $\theta(\infty, k^{\prime},
k=2)=1-\lambda \exp(-2 \lambda)$. This result is the same as for
the point impurity case. At $k=2$, the minimum value of the jamming
limit is $\theta_{\rm min}(\infty, k^{\prime}, k=2)=1-e^{-1}/2$
at $\rho_o=k^{\prime}/(1+k^{\prime})$.
The minimum value does not change for the length of the impurity.

Using these analytical results for lattice substrates we can obtain the
coverage for the continuum case. In the continuum limit objects of unit
length are deposited on a lattice initially occupied by 
impurities. 
Let the initial density of the impurities be $\mu$ in the continuum
limit. Rescale the density according to $k\rho_o = \mu$ and the time as
$kt=\tau$\cite{BK}.  With the rescaled density and time
remaining finite, we take the limit $k \rightarrow \infty$ of Eq.(18)
and Eq.(22).  When we take the continuum limit, the $k \rightarrow
\infty $ limit is primary. 
When the impurities are distributed randomly and
sequentially we use Eq.(18). For the case of point impurities 
the continuum coverage was already discussed by Ben-Naim and Krapivsky
\cite{BK}. When $k^{\prime} =2$, $t_o = - \ln [ 1+\frac{1}{2} \ln(1-
\rho_o) ] $. The continuum coverage is obtained as 
\be
\theta(\tau) = \exp(-\mu / 2) \int^{\tau}_0 dv \exp \left[
-2 \int^{v+\mu /2}_{\mu /2} dw \frac{1-e^{-w}}{w} \right]
\ee
In the limit $\mu \rightarrow 0$, the coverage converges to the
R\'{e}ni number $\theta(\infty)=R=0.7475 \cdots $ \cite{BK,Reni}.
In the limit $\mu \rightarrow \infty$, the coverage approaches zero
exponentially  according to $\theta(\infty) = (\mu /2) \exp(-\mu /2)$.
When $k^{\prime} >2 $, it is difficult to obtain the explicit
dependence of the initial time $t_o$ on $\rho_o$. Thus, for $k^{\prime}
\rightarrow \infty$ and $k \rightarrow \infty$ wih $k^{\prime}/k$
finite, we can not derive the general expression 
for the continuum limit when $k^{\prime} > 2$.

When the impurities are randomly distributed, we use Eq.(22). At
$k^{\prime} =2$, the continuum coverage is the same as Eq.(23).
The general form of the continuum limit in the case of 
random initial conditions is derived by the methods of
continuous RSA (not included the detailed calculations).
When $k^{\prime} \rightarrow \infty$ and $k \rightarrow \infty$ with
$k^{\prime}/k=l$ finite, we can obtain the continuum limit coverage 
as
\be
\theta(\infty) =\rho_o +(1-\rho_o) \exp(- \alpha)
\int_{0}^{\infty} dt \exp \left[ -2 \int_{\alpha}^{\alpha+t} du
\frac{1-e^{-u}}{u} \right]
\ee
where $\alpha=\rho_o / [ (1-\rho_o)l]$. When $\rho_o=0$, continuous RSA
is recovered. When $l \rightarrow 0$ and $\rho_o \rightarrow 0$ such
that $\rho_o / l = \mu$=const, then Eq.(23) is recovered. When $\rho_o
\rightarrow 1$ the coverage is only slightly higher than the initial
coverage, $\theta(\infty) = \rho_o + (\rho_o / l) \exp(-\alpha)$. 
In the continuum limit the coverage follows
the algebraic decay $\theta(\infty) -\theta(t) \sim t^{-1}$.

In summary we calculated the jamming limits for $k$-mers
on one-dimensional substrates 
for the random sequential adsorption initial condition and for the
random initial condition,
 by solving the appropriate rate equations.
The jamming limits $\theta(\infty;k^{'},k)$
show a minimum value at a particular density of impurities. The Monte
Carlo data are in good agreement with the analytical results.
The coverage in the continuum limit was discussed using the analytical
results for the lattice models.

\acknowledgments
This work was supported by Inha University, 1995 and by
the Basic Science Institute Program, Ministry of Education, 1996, 
Project No. BSRI-96-2430. I wish to thank professor Vladimir Privman 
for his careful reading of this manuscript. I am grateful to 
a referee who introduce the case for random initial condition.

\newcommand{\jpa}{J.Phys.A:Math.Gen.}

\begin{figure}
\noindent
\caption{ The jamming limits $\theta(\infty; k, k^{'})$ versus the
concentration of $k^{'}=2$ impurities $\rho$ for $k=3 (\bullet),
4(\circ)$ and $8(\Box)$. The symbols are Monte Carlo results and the
lines are analytical results.}
\label{fig1}
\end{figure}
\end{document}